\documentclass[twocolumn, unsortedaddress, superscriptaddress]{revtex4}
\usepackage{graphicx}
\usepackage{color}
\usepackage{amsmath}

\newcommand{\be}{\begin{equation}}
\newcommand{\ee}{\end{equation}}
\newcommand{\bea}{\begin{eqnarray}}
\newcommand{\eea}{\end{eqnarray}}

\begin{document}

\title{Hot-carrier photocurrent effects at graphene-metal interfaces.}

\author{K.J. Tielrooij}\thanks{Equal contribution} \affiliation{ICFO - Institut de Ci\'encies Fot\'oniques, Mediterranean Technology Park, Castelldefels (Barcelona) 08860, Spain}
\author{M. Massicotte}\thanks{Equal contribution}\affiliation{ICFO - Institut de Ci\'encies Fot\'oniques, Mediterranean Technology Park, Castelldefels (Barcelona) 08860, Spain}
\author{L. Piatkowski}\affiliation{ICFO - Institut de Ci\'encies Fot\'oniques, Mediterranean Technology Park, Castelldefels (Barcelona) 08860, Spain}
 \author{A. Woessner} \affiliation{ICFO - Institut de Ci\'encies Fot\'oniques, Mediterranean Technology Park, Castelldefels (Barcelona) 08860, Spain}
\author{Q. Ma} \affiliation{Department of Physics, Massachusetts Institute of Technology, Cambridge, MA 02139, USA}
\author{P. Jarillo-Herrero} \affiliation{Department of Physics, Massachusetts Institute of Technology, Cambridge, MA 02139, USA}
\author{N.F. van Hulst}\affiliation{ICFO - Institut de Ci\'encies Fot\'oniques,
Mediterranean Technology Park, Castelldefels (Barcelona) 08860,
Spain}\affiliation{ICREA-Instituci\'o
Catalana de Recerca i Estudis Avan\c{c}ats, Passeig Llu\'{\i}s Companys,
23, 08010 Barcelona, Spain}
\author{F.H.L. Koppens} \affiliation{ICFO - Institut de Ci\'encies Fot\'oniques,
Mediterranean Technology Park, Castelldefels (Barcelona) 08860,
Spain}

\newpage

\begin{abstract}

Photoexcitation of graphene leads to an interesting sequence of phenomena, some of which can be exploited in optoelectronic devices based on graphene. In particular, the efficient and ultrafast generation of an electron distribution with an elevated electron temperature and the concomitant generation of a photo-thermoelectric voltage at symmetry-breaking interfaces is of interest for photosensing and light harvesting. Here, we experimentally study the generated photocurrent at the graphene-metal interface, focusing on the time-resolved photocurrent, the effects of photon energy, Fermi energy and light polarization. We show that a single framework based on photo-thermoelectric photocurrent generation explains all experimental results.

\end{abstract}

\maketitle

\section{Introduction}

Graphene photodetectors and light harvesting devices benefit from graphene's unique optical properties with extremely broadband, wavelength-independent absorption from the ultraviolet to the far-infrared, and its outstanding electrical properties with high mobilities and gate-tunable carrier densities \cite{Bonaccorso2010}. These optoelectronic properties are supplemented by mechanical flexibility and strength, and the potential to be integrated with existing technologies. As a result, graphene has already been used for demonstrations of a number of promising optoelectronic devices \cite{Koppens2014}.
\\

Of special interest are devices where light is converted into an electrical signal, i.e\ a photovoltage and/or photocurrent. One of the simplest device geometries for generating photocurrent in graphene is a graphene sheet contacted by two metal contacts that serve as source and drain. The Fermi energy $E_F$ of the graphene sheet can be controlled through capacitive coupling using a doped silicon back gate, separated from the graphene sheet by an oxide. In these devices, photocurrent is generated when light is focused at the interface between graphene and the metal contacts. The fabrication of such devices is not demanding and relatively easy to scale up for commercial production. The photocurrent generated at the graphene-metal interface has been studied since 2008 \cite{Park2009, Lee2008,Xia2009, Mueller2009} and has been shown to give rise to an ultrafast photoresponse with picosecond switching dynamics \cite{Urich2011, Sun2012, Xia2009}. The response is furthermore extremely broadband, covering the visible, infrared and far-infrared (THz) wavelength ranges \cite{Cai2014,Zhang2013, Koppens2014}. The photocurrent can moreover be enhanced by plasmonic effects \cite{Echtermeyer2011} and by suspending the graphene sheet \cite{Freitag2013}. It has also been shown that the gate-response of the photocurrent depends on the choice of the metal for the contacts \cite{Lee2008,Mueller2009} and the light polarization \cite{Echtermeyer2014, Kim2012}. These intriguing aspects have so far not been explained within a single framework.
\\

Here, we study photocurrent at the graphene-metal interface and explain the experimental results within one general framework of photo-thermoelectric (PTE) current generation. The PTE effect has been shown to be the dominant photocurrent generation mechanism at graphene $pn$-junctions \cite{Gabor2011, Song2011, Graham2013} and at interfaces of single layer and bilayer graphene \cite{Xu2010}. We show that using the PTE framework we can explain: the time-resolved photocurrent dynamics that we observe at the graphene-metal interface (Section IV), the dependence of the photocurrent on the photon energy and the type of substrate (Section V), the effect of the Fermi energy and of the type of metal used for the contacts on the photocurrent (Section VI), and finally the influence of the polarization of the incoming light on the generated photocurrent (Section VII). These results are useful for assessing the potential and limitations of device performance parameters, such as the photodetection speed, photoconversion efficiency, and spectral response, among others.
\\

\section{The photo-thermoelectric effect in graphene}

The photo-thermoelectric generation of photocurrent is based on the thermoelectric effect, where a temperature gradient $\nabla T$ is directly converted into a voltage $V_{\rm TE}$ that is generated by the diffusion of charge carriers from the hot to the cold region. This process is governed by the Seebeck coefficient that is defined as $S = V_{\rm TE}/\nabla T$. In graphene, the Seebeck coefficient is typically much larger than that of, for instance,  gold \cite{Wei2009}. It is furthermore tunable through the Fermi energy $E_F = k_B T_F$, with $k_B$ Boltzmann's constant and $T_F$ the Fermi temperature. This is the case because the hot and cold regions correspond to different Fermi-Dirac distributions (see Fig.\ 1a), which are determined by $E_F$ and electron temperature $T_{\rm el}$. The thermally induced charge diffusion in graphene thus depends on these Fermi-Dirac distributions, and also on the energy-dependent scattering time $\tau (\epsilon)$ of the electrons, where hotter electrons could be more/less mobile. In the case of relatively high Fermi energy ($T_F > T_{\rm el}$), the Seebeck coefficient is then given by \cite{Hwang2009}
\be S = {{2\pi^2 k_B T_{\rm el}}\over{3 e T_F }} \hspace {2mm}. \ee
This assumes charged impurity scattering as the dominant process, which corresponds to a scattering time that scales linearly with electron energy, i.e.\ $E_F$-independent mobility \cite{dassarmareview}. For a typical Fermi energy of 0.1 eV, this gives a room temperature Seebeck coefficient of $\sim$0.1 mV/K.
\\

In the photo-thermoelectric effect, the temperature difference is created by photoexcitation. Absorbed photons in graphene lead to ultrafast \cite{Breusing2009, Breusing2011} and efficient \cite{Winzer2010, Tielrooij2013, SongPRB2013, Jensen2014} carrier heating. The electron distribution after photoexcitation is characterized by an elevated 'hot' electron temperature $T_{\rm el,hot}$, compared to the electron temperature without photoexcitation $T_{\rm el,0}$ (see Fig.\ 1a). After local photoexcitation and local carrier heating, diffusion occurs between the photoexcited 'hot' region and the region without photoexcitation, governed by the Seebeck coefficient $S$. If a homogeneous graphene sheet is locally photoexcited, this leads to radial charge carrier diffusion, where no net photocurrent is generated due to the isotropic charge current density. An anisotropic charge current density is created when an interface between regions of different Seebeck coefficients, $S_1$ and $S_2$, is photoexcited (see Fig. 1a). This is the case at the interface of single layer and bilayer graphene \cite{Xu2010} and at the interface of graphene with different Fermi energies \cite{Song2011, Gabor2011, Freitag2013, Graham2013}. The generated PTE photovoltage is then given by
\be V_{\rm PTE} = (S_2 - S_1)(T_{\rm el,hot} - T_{\rm el,0}) \hspace{2mm}. \ee
The PTE photovoltage generation process benefits from absorbed photon energy being converted efficiently into heat in the electron system, rather than into lattice heat \cite{Tielrooij2013, Jensen2014, SongPRB2013, Song2014}. This, in combination with the small electron heat capacity, compared to the phonon heat capacity, means that the electrons can reach a temperature easily exceeding 1000 K for a photon fluence on the order of a $\mu$J/cm$^2$ \cite{Lui2010, Breusing2011, Tielrooij2013, Tielrooij2014, Gierz2013, Johannsen2013}. This high electron temperature together with the considerable Seebeck coefficient of graphene leads to a substantial PTE photovoltage in graphene devices.
\\

\section{Devices}

To study photocurrent generation at the interface between graphene and a metal contact, we use three different samples that contain -- besides graphene-metal interfaces -- other interfaces, where the photocurrent generation mechanism has been established to be dominated by the photo-thermoelectric effect. This enables us to compare the PTE response at these interfaces to the response at the graphene-metal interface.
\\

The first device is a \textit{dual-gated device} that consists of graphene on a substrate with a doped silicon back gate (separated from the graphene sheet by 300 nm of SiO$_2$) and a local metal top gate (separated by 10-20 nm of hexagonal boron nitride) that can both be used to change the Fermi energy of the exfoliated graphene flake (see Fig.\ 1b). At the interface between the region that has a Fermi energy determined by the back gate and the region with a Fermi energy determined by the top gate, the photocurrent is dominated by the photo-thermoelectric effect, as demonstrated theoretically \cite{Song2011} and experimentally \cite{Gabor2011, Graham2013}. This device also contains two metal contacts and thus two graphene-metal interfaces. More details on the fabrication of this device can be found in Ref.\ \cite{Gabor2011}.
\\

Our second device is a \textit{globally gated device} in the most common field-effect transistor geometry (see Fig.\ 1c). This device contains an exfoliated graphene flake, with two metal contacts. The back gate is formed by doped silicon, separated from the graphene sheet by 285 nm of SiO$_2$. Photocurrent is created when light is focused at the interface of graphene and the metal contacts.
\\

Finally, the third device is a \textit{transparent substrate device}, with an exfoliated flake on top of a substrate that consists of only SiO$_2$ (see Fig.\ 1d). The Fermi energy of this device is not tunable. However, the device offers three different interfaces to study photocurrent generation: a graphene-metal interface, an interface of single layer graphene (SLG) and bilayer graphene (BLG) and an interface between bilayer graphene and multilayer graphene. In the case of the SLG-BLG interface, the photocurrent mechanism has been established to be PTE \cite{Xu2010}.
\\

\section{Time-resolved photocurrent}

The generation of a PTE voltage after photoexcitation is closely connected to the heating and cooling dynamics of the electron system. The time scale of the heating process, for instance, determines the heating efficiency \cite{Jensen2014}. Furthermore, the generated photovoltage $V_{\rm PTE}$ only exists as long as the hot electron distribution exists, which means that the time-averaged, steady-state photovoltage $\overline{V}_{\rm PTE}$ depends on the life time of the hot electrons: $\overline{V}_{\rm PTE} \propto 1/\Gamma_{\rm cool}$ (for laser spot size larger than cooling length), with $\Gamma_{\rm cool}$ the cooling rate. So a lower cooling rate (longer lifetime of hot electrons) leads to a larger photocurrent. The electron heating and cooling dynamics in bulk graphene have been studied using pump - probe measurements, such as optical pump - probe \cite{Ploetzing2014, Breusing2011, Brida2013, Lui2010}, femtosecond time-resolved angle-resolved photo-electron spectroscopy (ARPES) \cite{Gierz2013, Johannsen2013}, and time-resolved optical pump - terahertz (THz) probe spectroscopy \cite{Jnawali2013,Docherty2012,Frenzel2013, Frenzel2014, Strait2011, George2008, Tielrooij2013, Shi2014}. The photoexcited carrier dynamics have also been studied in graphene-based devices through time-resolved photocurrent scanning microscopy \cite{Graham2013, Sun2012, Urich2011, Tielrooij2014}.
\\

These studies indicate that light absorption leads to the following dynamics (see Ref.\ \cite{Song2014} and references therein for a more detailed treatment): Absorbed light induces electron-hole pair excitation, assuming that the photon energy $E_{\rm exc}$ is more than twice as large as the Fermi energy $E_F$. This creates a non-equilibrium state with very hot electrons at an energy $E_{\rm exc}/2$. This is followed by ultrafast ($<$50 fs) electron heating, which creates a quasi-equilibrium distribution that can be described by an increased electron temperature. The details of this heating process have been addressed in a number of experimental \cite{Gierz2013, Johannsen2013, Breusing2009, Breusing2011, Brida2013, Tielrooij2013, Jensen2014, Lui2010, Tielrooij2014} and theoretical \cite{Winzer2012, Winzer2010, Tomadin2013, SongPRB2013, Song2014} studies. The system returns to its original (pre-photoexcitation) state through cooling of the hot electrons, which can occur through interaction with graphene lattice optical or acoustic phonons, and substrate phonons \cite{Freitag2013, Alencar2014, Ma2014, Graham2013, Song2012, Song2014}. At room temperature, disorder-assisted supercollisions with energy transfer to acoustic phonons were found to dominate the cooling process \cite{Song2012, Song2014, Ma2014, Graham2013}.
\\

These electron temperature dynamics have been studied in quite some detail at $pn$-junctions \cite{Sun2012,Graham2013,Tielrooij2014}. To establish a better understanding of the mechanism and dynamics of the photoresponse near contacts, we compare the photovoltage dynamics for the two regions (at the contact and at the $pn$-junction). We apply ultrafast time-resolved photocurrent scanning microscopy measurements to our \textit{dual-gated device} and compare the dynamics at the $pn$-junction with the dynamics at the graphene-metal interface. The setup is very similar as the ones described in Refs.\ \cite{Tielrooij2014, Sun2012, Graham2013, Urich2011} and uses pulse pair excitation with two ultrashort pulses (with a wavelength of $\sim$800 nm) and a variable time delay between the two pulses. Due to an intrinsic nonlinearity (the electron heat capacity of graphene depends on electron temperature \cite{Kittel}), a lower photocurrent is generated when the two pulses overlap in time, than when they contribute to photocurrent independently, i.e.\ when the time delay is larger than the carrier cooling time. We refer to the lowering of the photocurrent at short time delays as the photocurrent dip. The dynamics of the photocurrent dip directly reflect the temperature dynamics of the photoexcited electrons in graphene \cite{Tielrooij2014, Sun2012, Graham2013, Urich2011}.
\\

Figure 2a shows the results for the PTE photocurrent that is generated at the $pn$-junction, together with a numerical calculation of the delay-time dependent photocurrent dip (see Refs.\ \cite{Graham2013, Tielrooij2014} for details). The dynamics correspond to a photocurrent generation time $<$200 fs (our time resolution in this experiment) and a relaxation time of 1.4 ps. The time-resolved photocurrent measurements on the same device, under the same conditions, but with the laser focused at the graphene-metal interface is shown in Fig.\ 2b. These dynamics, with a photocurrent generation time below 200 fs and a relaxation time of 1.2 ps, are strikingly similar to the dynamics at the $pn$-junction. For both the $pn$-junction and the graphene-metal interface these dynamics are in agreement with photocurrent generation corresponding to femtosecond carrier heating \cite{Breusing2011, Gierz2013, SongPRB2013}, and relaxation corresponding to picosecond supercollision cooling \cite{Song2012, Song2014, Ma2014, Graham2013}. From these data we conclude that the cooling dynamics near the contact likely have the same origin as at the $pn$-junction, namely supercollision cooling \cite{Song2012, Song2014}.
\\

\section{Spectrally resolved photocurrent}

The spectral response of photosensing and photovoltaic optoelectronic devices is an important device characteristic. The dependence of the electron temperature, and thus the photocurrent, on photon energy is, for instance, strongly related to the carrier heating efficiency \cite{Tielrooij2013}, a crucial parameter for PTE-based devices, since it is directly linked to the device sensitivity. Here, we examine the spectrally resolved photoresponse using a photocurrent scanning microscopy setup with a variable excitation wavelength in the range 500 -- 1500 nm (2.5 -- 0.8 eV) (see inset of Fig.\ 3a). We measure the external responsivity $R_{\rm ext} = I_{\rm PC}/P_{\rm exc}$, with $I_{\rm PC}$ the photocurrent and $P_{\rm exc}$ the excitation power. Again, we compare the response at the $pn$-junction, which has been studied in detail in Ref.\ \cite{Tielrooij2014}, with the response at the graphene-metal interface.
\\

The photoresponse for the \textit{dual-gated device} at the $pn$-junction is shown in Fig.\ 3a and at the graphene-metal interface in Fig.\ 3b. The inset in Fig.\ 3c schematically shows the spectrally-resolved measurement technique. The photoresponse at the $pn$-junction is wavelength-dependent in a non-monotonous fashion. The reason for this response is that the graphene absorption $\alpha (\lambda)$ depends on wavelength $\lambda$ due to reflections at the Si--SiO$_2$ interface \cite{Tielrooij2014} -- an effect that is very similar to the one that makes graphene visible when using a similar substrate \cite{Blake2007}. Indeed, the photocurrent has a very similar wavelength dependence as the absorption that was calculated using numerical software (Lumerical FDTD Solutions software), taking into account the multilayer substrate. The external photoresponsivitity is thus wavelength-dependent, as a result of the substrate that is used.
\\

In Fig.\ 3c we show the ratio between the photocurrent generated by focusing the laser at the contact over the photocurrent generated with the laser at the $pn$-junction for the \textit{dual-gated device}. This divides out the substrate-induced wavelength-dependent absorption. It has furthermore been established that the $pn$-junction gives a PTE response with a flat intrinsic (absorption-corrected) responsivity \cite{Tielrooij2014}. This means that Fig.\ 3c directly reflects the intrinsic wavelength dependence of the photocurrent generated at the contact. We find that the response is almost flat above $\sim$600 nm and increases below 600 nm. These observations lead to important conclusions on the photocurrent mechanism and the conversion efficiency of absorbed light into hot electrons. A flat photoresponsivity means that the photocurrent is wavelength-independent for constant power. However, constant power corresponds to fewer photons at higher photon energy, which means that a high energy photon leads to a proportionally higher electron temperature than a low energy photon. This is in strong contrast to photovoltaic devices, where the photoresponse is determined by the absorbed photon flux, giving a lower (power-normalized) responsivity at shorter wavelengths \cite{Sze}. From the flat response in Fig.\ 5c above 600 nm, we thus conclude that the PTE effect is the dominant photocurrent generation mechanism at the graphene-metal interface.
\\

The microscopic picture that explains why a higher energy photon gives a larger photoresponse is that a higher energy photon leads to a proportionally larger number of intraband energy scattering events, which in turn lead to a higher electron temperature and thus a larger photovoltage \cite{SongPRB2013, Song2014, Tielrooij2014}. Terahertz photoconductivity measurements, which also probe the electron temperature, found a similar linear scaling with photon energy \cite{Tielrooij2013}. It was furthermore shown that a linear relation between electron temperature and photon energy corresponds to highly efficient carrier heating \cite{Tielrooij2013,Tielrooij2014}. The reason for the efficient heating is the ultrafast timescale associated with this process, which dominates over alternative energy relaxation pathways, such as acoustic and optical phonon emission, provided that the electron temperature is below 3000 K (i.e.\ for $k_BT_{\rm el} <$ optical phonon energy) \cite{Jensen2014}. The wavelength-independent $internal$ responsivity means that the photon-flux-normalized response increases linearly with photon energy, which shows that the energy transfer from absorbed photons to hot electrons is efficient for both $pn$-junction \cite{Tielrooij2014} and at the graphene-metal contact.
\\


We observe that below $\sim$600 nm the photocurrent ratio increases quite strongly. Interestingly, this increase corresponds well with the wavelength-dependent absorption of the gold contacts, calculated using the complex refractive index of gold from Ref.\ \cite{Rakic1998}. This correspondence was also observed in Ref.\ \cite{Echtermeyer2014} and can be understood by taking into account the contact-heating-induced thermoelectric effect: absorbed light and subsequent heat dissipation in a gold contact lead to local heating of the graphene sheet, generating a photocurrent. An analogous effect was observed recently by resonantly exciting SiO$_2$ substrate phonons with mid-infrared light, which also leads to photocurrent enhancement \cite{Badioli2014, Freitag2014}. Thus, the photocurrent at the graphene--metal interface is a combination of the PTE effect due to light absorption in graphene and the thermoelectric effect due to light absorption in the metal contacts, where the former dominates above 600 nm. In the case of photocurrent generation that is induced by light absorption in the contacts, the photocurrent is not only generated at the graphene--metal interface. Rather, the photocurrent extends spatially into the contacts, as shown in Figs.\ 3d-e. Indeed, this occurs mainly for wavelengths that correspond to significant gold absorption.
\\

%

\section{Gate-dependent photocurrent}


Due to the $E_F$-dependent Seebeck coefficient, the PTE photocurrent response is strongly gate-tunable \cite{Lemme2011}, which is an interesting feature for optoelectronic devices that would require an electrically controllable photoresponse. We now examine the gate-response at the $pn$-junction of the \textit{dual-gated device} and at the metal-graphene interface of the \textit{globally gated device} and evaluate the results within the framework of PTE photovoltage creation. In Fig.\ 4a we show that the \textit{dual-gated device} at the $pn$-junction shows two sign changes as a function of back gate voltage (with an excitation wavelength of 800 nm and a top gate voltage of 0.4 V, so that the graphene region whose carrier density is determined by the top gate, is tuned away from the Dirac point). It has been shown that PTE photocurrent in such a device should indeed result in two sign changes: one when the two chemical potentials are equal and another one when the graphene whose carrier density is determined by the back gate, is tuned through the Dirac point  \cite{Gabor2011, Song2011, Graham2013}.
\\

The photocurrent for the \textit{globally gated device} as a function of back gate voltage (for 630 nm excitation) shows a symmetric signal with a sign change around the Dirac point (see Fig.\ 4b), similar to what has been reported earlier \cite{Lee2008}. While the double sign change is a clear signature of the PTE effect, we now argue that PTE can also give rise to a symmetrical gate response near the contacts. In the most simple approach, the photovoltage is given by $V_{\rm PTE} = (S_2 - S_1)(T_{\rm el,hot} - T_{\rm el,0}) $. If $S_2$ represents the Seebeck coefficient of the graphene underneath the metal contact and there would be very little metal-induced doping, the photocurrent would only depend on the gate dependence of $S_1$. This then leads to a symmetric gate response as in Fig.\ 4b, assuming that photoexcitation is similar in both graphene regions outside and underneath the metal contact.
\\

In a more realistic approach, where we take into account the high reflectivity of the metal contact, we numerically simulate the PTE photocurrent response at the graphene--metal interface using a spatial profile of the Seebeck coefficient $S (x)$ and a spatial profile of the electron temperature $T_{\rm el} (x)$. This will generate a local photovoltage $V_{\rm PTE} = \int dx S(x) \nabla T_{\rm el} (x)$. For the Seebeck coefficient profile we use three regions: the first region corresponds to graphene underneath the gold contact, with a Fermi energy that is pinned at $S_{\rm g}' = 5$ $\mu$V/K or at $S_{\rm g}'' = -30$ $\mu$V/K; the next region is a transition region between graphene that is pinned by the contact and the gate-tunable graphene sheet; and finally we have the gate-tunable graphene region with $S_{\rm g} (V_{\rm bg})$. For the spatial profile of the hot electrons we take into account the Gaussian beam profile of the laser focus and the strong reflection of incident light at the gold contact. We note that the width of the Seebeck regions, their numerical values, and the shape of the hot electron profile do not influence the qualitative shape of the gate-dependent PTE photocurrent. However, it is essential to include the transition region between pinned graphene underneath the metal and gate-tunable graphene. Using $S (x)$ and $T_{\rm el} (x)$ we find the gate-dependent photocurrent traces in Fig.\ 4d. This reproduces the symmetric dependence with a sign change close to the Dirac point, for the case of low metal-induced doping ($S_{\rm g}'$). By changing the metal-induced doping of the graphene underneath the contact to $S_{\rm g}''$ (see Fig.\ 4c-d), we can also create a less symmetric gate response, with a sign change that occurs at a higher or lower voltage than the voltage that corresponds to the Dirac point, as observed for instance in Refs.\ \cite{Lee2008, Mueller2009}. Whereas this model reproduces the experimentally observed trends, it merely serves as an example to demonstrate that the observations can be explained by PTE-generated photocurrent at the graphene-metal contact.
\\

\section{Polarization-resolved photocurrent}

We complete our study of the PTE photocurrent generated at graphene-metal interfaces by investigating its dependence on the polarization of the incident light, an experimental variable which is known to influence the dynamics of photoexcited charges. It has indeed been predicted \cite{Gruneis2003} and shown experimentally \cite{Mittendorff2014} that linearly polarized light generates a very short-lived ($\sim$150 fs) anisotropic carrier distribution in momentum space. We now investigate the effect of this anisotropy on the photoresponse. In Fig.\ 5a we compare the polarization dependence of the photocurrent generated at the three different interfaces of the \textit{transparent substrate device}. The light polarization appears to have no effect on the photocurrent at the SLG-BLG and BLG-graphite interfaces, from which we conclude that the initial anisotropic carrier distribution directly after photon absorption does not affect the photocurrent magnitude. The reason for this is that the PTE photocurrent response depends on the temperature of the carrier distribution, rather than on its momentum distribution. Furthermore, the PTE photocurrent is generated during the time interval that carriers are hot, which is 1-2 picoseconds (see Section IV) and thus much longer than the lifetime of the anisotropic carrier distribution. Therefore, light polarization does not have an effect on the intrinsic PTE response.
\\

In contrast, the photocurrent at the graphene-metal interface for 630 nm excitation displays a strong dependence on polarization, with a maximum (minimum) photocurrent when the polarization is perpendicular (parallel) to the metal contact edge.  We observe this effect at the graphene-metal interface of every one of the $\sim$10 devices that we have measured. This effect is reduced for excitation with 1500 nm light, compared to 630 nm excitation. Figure 5b shows a polarization-resolved photocurrent map of the metal-graphene interface (of the \textit{globally gated device}) which is obtained by measuring the photocurrent as a function of polarization at many different positions (630 nm excitation). This vector map clearly reveals that the photocurrent is enhanced when light polarization is perpendicular to the contact edge. A similar effect was observed in Ref.\ \cite{Echtermeyer2014}, whereas Ref.\ \cite{Kim2012} reports the opposite effect, i.e.\ a maximum photocurrent for polarization parallel to the metal contact edge.
\\

The observation of a polarization-dependent photocurrent at the graphene-metal interface, together with the absence of polarization effect at the SLG-BLG interface, suggests that an extrinsic factor affects the photoresponse at the graphene-metal interface. The extrinsic factor we consider is the effect of the metal contacts on the electrical field intensity and thereby the light absorption in the graphene sheet. We perform simulations using a 2D Maxwell equations solver (Lumerical FDTD Solutions software) for 630 nm and 1500 nm excitation, and find that for perpendicular polarization (with respect to the metal contact edge) the electric field is enhanced and confined at the graphene-metal interface (see inset of Fig.\ 5c). This is a phenomenon known in photonics as the lightning-rod effect. Due to this photonic effect, the energy absorbed by graphene close to a metal edge varies with polarization $\alpha (\angle)$, reaching a maximum when the polarization is perpendicular to the contact edge $\alpha (\perp)$. In Fig.\ 5c we show the normalized light absorption in the graphene sheet with and without the presence of a gold contact (for 630 and 1500 nm light). We observe no polarization contrast when there is no gold contact, whereas the presence of a contact leads to polarization contrast, which is stronger for 630 nm excitation than for 1500 nm excitation. Thus, we observe very similar behavior for the light absorption in graphene (Fig.\ 5c) and for  the photocurrent at the graphene-metal contact (Fig.\ 5a) as a function of polarization. These similarities arise, because the PTE photocurrent depends on the power absorbed in graphene, which is subsequently converted into electron heat. A polarization-dependent graphene absorption $\alpha (\angle)$ will therefore give rise to a corresponding dependence of the PTE photocurrent.
\\

Interestingly, the polarization contrast in some cases depends on the gate voltage, which is shown in Fig.\ 5d. The most dramatic polarization contrast is observed near the Dirac point, where even the sign of the photocurrent changes with polarization. We explain these observations by taking into account two contributions to the photocurrent: (1) the PTE photocurrent generated by light absorption in graphene, which, as we demonstrated above, depends on polarization; and (2) the thermoelectric photocurrent originating from the absorption of light in the bulk gold contact. These two contributions not only have a different polarization dependence, but also a slightly different gate-response. Using a laser wavelength at which gold (weakly) absorbs (630 nm in this experiment), both contributions lead to a photoresponse. Then by changing the gate voltage, we are able to tune the relative contribution of each photocurrent contribution and reach a point where the sign of the photocurrent depends on polarization. This effect could be useful for applications such as polarization detectors.
\\

\section{Summary and conclusion}

Summarizing, our experimental results show that the PTE photocurrent that is generated at the graphene--metal interface exhibits \textit{i)} the same carrier dynamics as the PTE photocurrent at the $pn$-junction with $<$200 fs electron heating and $\sim$1-2 ps electron cooling (Fig.\ 2), \textit{ii)} a flat spectral response (above $\sim$600 nm) that shows PTE-dominated photocurrent generation and efficient electron heating (Fig.\ 3), \textit{iii)} a gate response that can be reproduced by a simple model based on the PTE effect, which also reproduces the effect of the metal used as contact material (Fig.\ 4), and \textit{iv)} a polarization response that depends on wavelength and gate voltage (Fig.\ 5). We furthermore find two photocurrent effects that are induced by the presence of a metal contact. The first effect of the metal contact concerns the absorption of light in the gold contact for excitation wavelengths below 600 nm, which leads to local heating and therefore an additional thermoelectric photocurrent \cite{Echtermeyer2014}. The second metal contact effect is a photonic effect that is associated with field confinement at the metal edge. This leads to polarization-dependent absorption $\alpha (\angle)$, which is maximum when the polarization of the light is perpendicular to the metal edge, and leads to enhanced photocurrent. We thus explain a wide range of different experimental results within one unifying framework of photo-thermoelectric photocurrent generation at the graphene-metal interface.
\\

\subsection*{Acknowledgements}
We thank Leonid Levitov and Justin Song for useful discussions. KJT thanks NWO for a Rubicon fellowship. MM thanks the Natural Sciences and Engineering Research Council of Canada (PGSD3-426325-2012). LP acknowledges financial support from Marie-Curie International Fellowship COFUND and ICFOnest program. FK acknowledges support by the Fundacio Cellex Barcelona, the ERC Career integration grant 294056 (GRANOP), the ERC starting grant 307806 (CarbonLight) and support by the E.\ C.\ under Graphene Flagship (contract no. CNECT-ICT-604391). NvH acknowledges support from ERC advanced grant ERC247330. QM and PJH have been supported by AFOSR Grant No. FA9550-11-1-0225 and a Packard Fellowship. This work made use of the Materials Research Science and Engineering Center Shared Experimental Facilities supported by the National Science Foundation (NSF) (Grant No. DMR-0819762) and of Harvard’s Center for Nanoscale Systems, supported by the NSF (Grant No. ECS-0335765).
\\

\clearpage
\onecolumngrid

\section{Figures}

\begin{figure} [h!!!!!]
   \centering
   \includegraphics [scale=0.6]
   {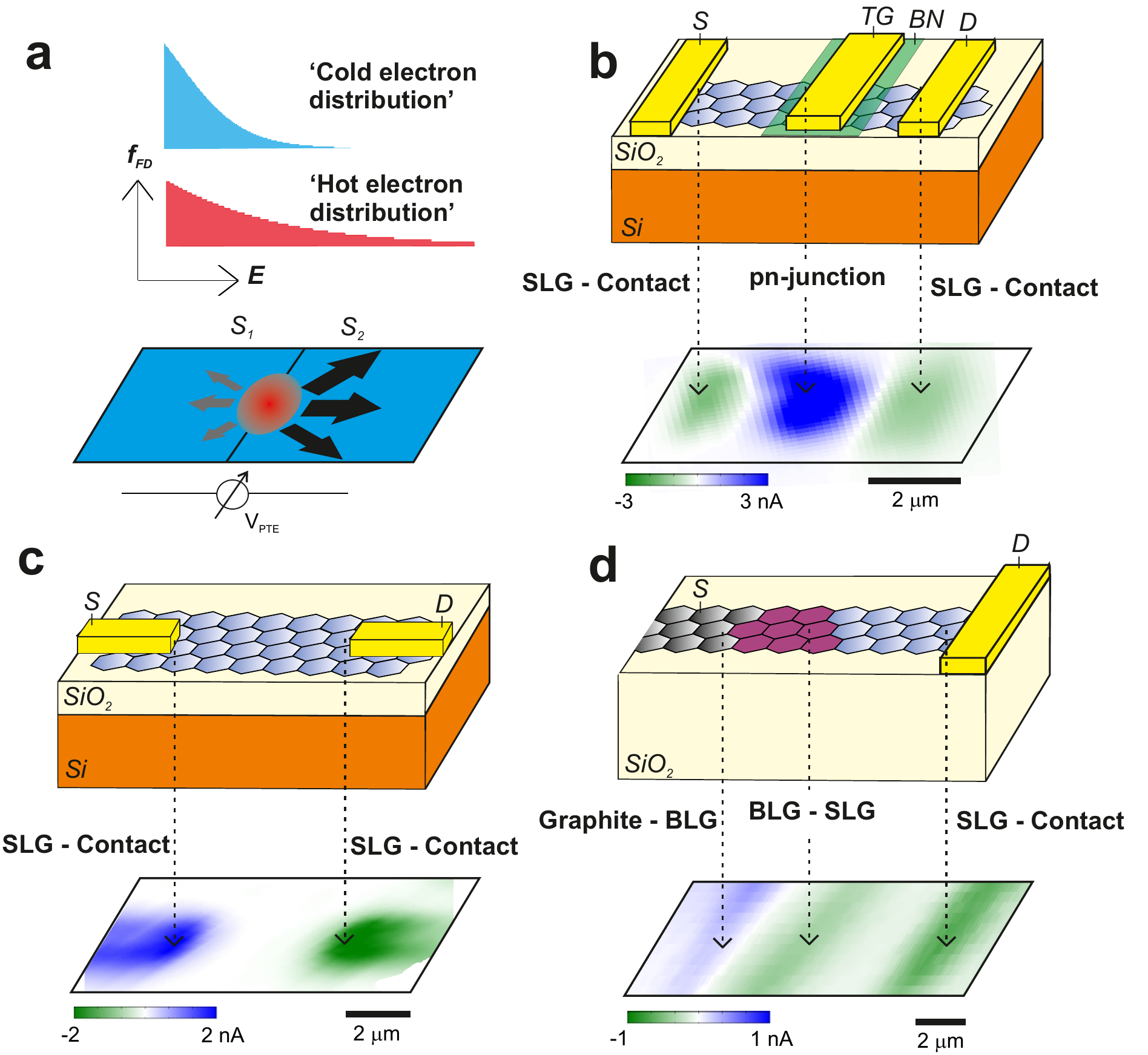}
\caption{\textit{Hot electron photocurrent and devices.}
\textbf{a)} Photo-induced electron heating in graphene leads to a broader Fermi-Dirac distribution (red), in comparison with the distribution without photoexcitation (blue). The carrier diffusion between photoexcited (''hot electron distribution'') and non-photoexcited (''cold electron distribution'') is governed by the Seebeck coefficient $S$. If hot electrons are created at an interface of two regions with different Seebeck coefficients $S_1$ and $S_2$, a net photo-thermoelectric voltage $V_{\rm PTE}$ is created due to net electron movement.
\textbf{b)} Device layout and photocurrent scanning microscopy image of the \textit{dual-gated device}, with a silicon back gate separated by 300 nm SiO$_2$, and a top gate (\textit{TG}) separated by hexagonal BN. The graphene (atomic structure not to scale) is contacted by source (\textit{S}) and drain (\textit{D}) contacts, through which photocurrent is measured.
\textbf{c)} Device layout and photocurrent scanning microscopy image of the \textit{globally gated device}, with a silicon back gate separated by 285 nm SiO$_2$ and graphene contacted by source (\textit{S}) and drain (\textit{D}) contacts.
\textbf{d)} Device layout and photocurrent scanning microscopy image of the \textit{transparent substrate device}, with a flake that contains adjacent regions of single layer graphene (SLG), bilayer graphene (BLG) and graphite.
}
    \label{Fig_PC_Maps}
\end{figure}

\clearpage

\begin{figure} [h!!!!!]
   \centering
   \includegraphics [scale=0.8]
   {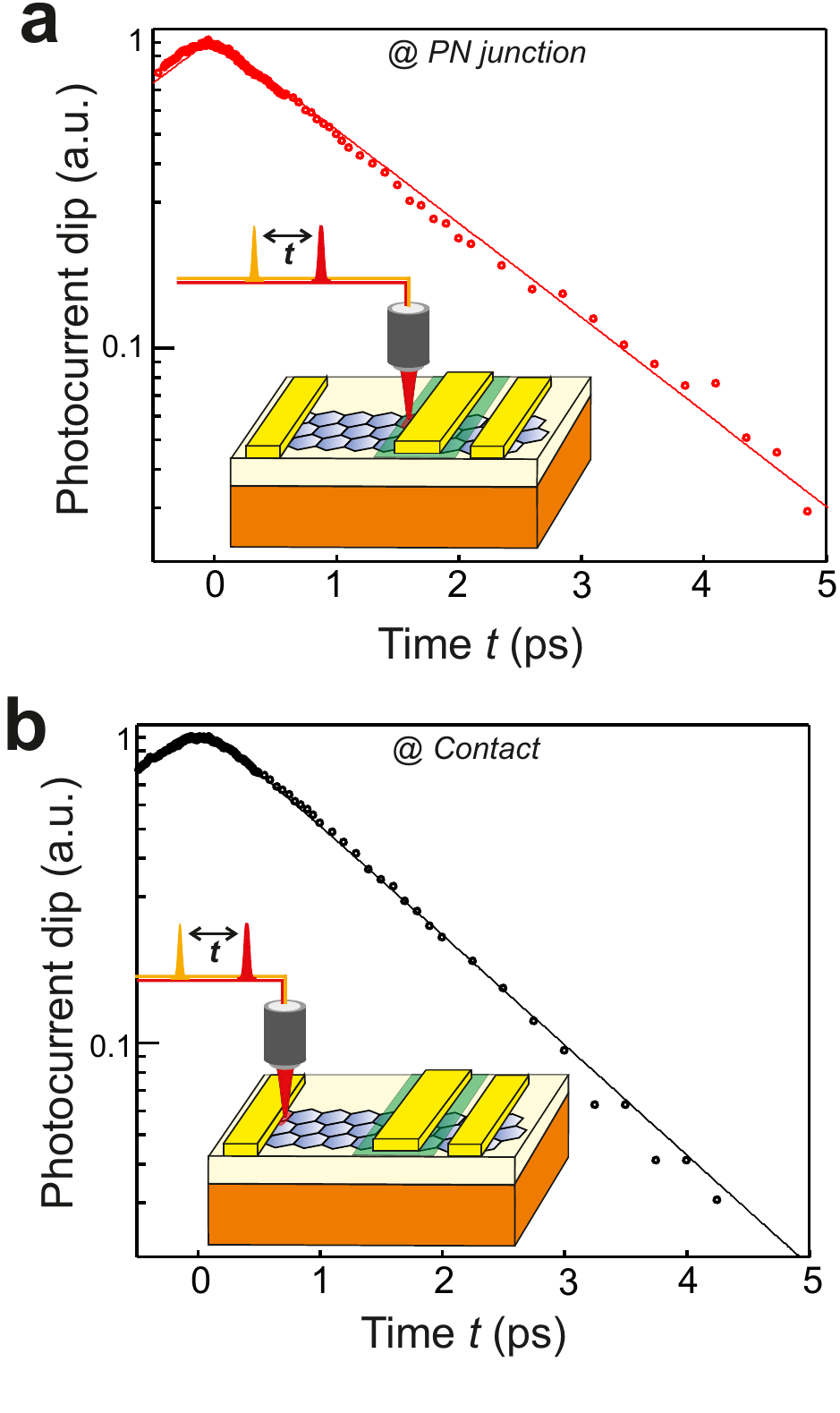}
\caption{\textit{Hot electron dynamics at $pn$-junction and graphene-metal contact.}
\textbf{a)} Experimental results (dots) of time-resolved photocurrent microscopy measurements at the $pn$-junction of the \textit{dual-gated device}, where a pulse pair with the pulses separated by a time $t$, create a dip in the photocurrent. The photon wavelength is 800 nm. The photocurrent dip as a function of delay time represents the electron temperature dynamics. The line describes the numerically calculated photocurrent dip, based on electron heating with a time scale $<$200 fs and an exponential cooling time of 1.4 ps. The inset shows the device and measurement configuration.
\textbf{b)} The experimental results (dots) of the same measurement as in \textbf{a}, now with the laser pulse-pair focused at the graphene-contact interface. The line describes the numerically calculated photocurrent dip, based on electron heating with a time scale $<$200 fs and an exponential cooling time of 1.2 ps.
}
    \label{Fig_TimeResolved}
\end{figure}

\clearpage

\begin{figure} [h!!!!!]
   \centering
   \includegraphics [scale=0.6]
   {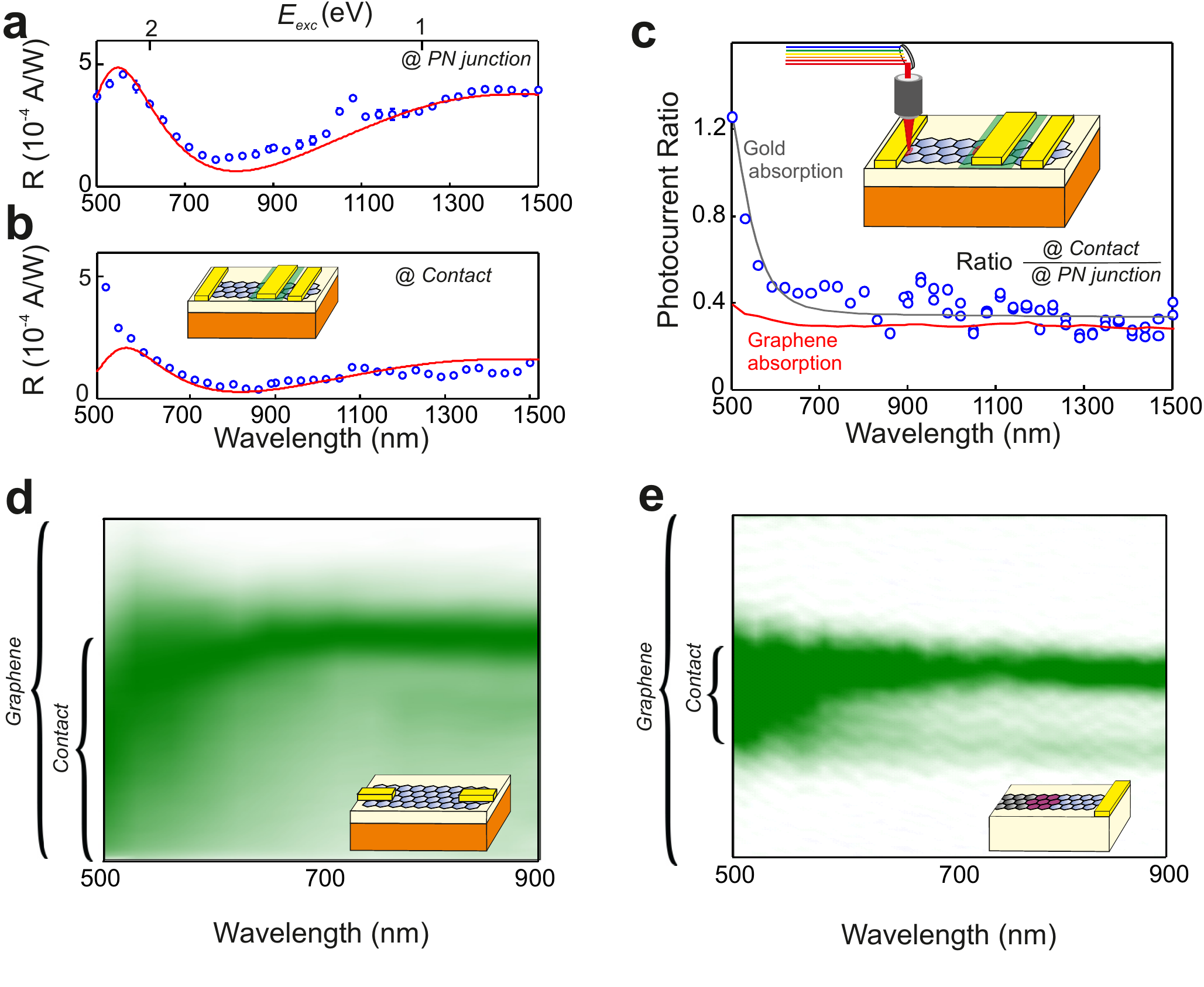}
\caption{\textit{Effect of photon energy on photocurrent.}
External responsivity with the laser focused at the $pn$-junction \textbf{(a)} and the graphene-metal interface \textbf{(b)} of the \textit{dual-gated device}, as a function of wavelength, together with numerical simulations of the wavelength-dependent absorption of the multilayer structure with reflections at the SiO$_2$-Si interface and a SiO$_2$ layer thickness of 300 nm (red line).
\textbf{c)} Ratio of the photocurrent with the laser focused at the graphene-contact interface divided by the photocurrent with the laser focused at the $pn$-junction for the \textit{dual-gated device}. The laser excitation has a power of $\sim$20 $\mu$W and a pulse duration of $\sim$30 ps. Above 600 nm, the ratio is almost flat, indicating that the spectral response at the contact is very similar to the spectral response at the $pn$-junction, where it is dominated by the PTE effect \cite{Gabor2011, Tielrooij2014}. Below 600 nm, the ratio increases, which suggests a sum of two photocurrent generation mechanisms: PTE photocurrent that is proportional to the graphene absorption (red line) and thermoelectric photocurrent due to gold heating, which is proportional to the gold absorption (grey line). The inset shows the measurement configuration, where the laser excites the device with a variable photon wavelength.
\textbf{d)} Spatially-resolved photocurrent as a function of wavelength for the \textit{globally gated device}.
\textbf{e)} Spatially-resolved photocurrent as a function of wavelength for the \textit{transparent substrate device}.
}
    \label{Fig_WavDep}
\end{figure}

\clearpage

\begin{figure} [h!!!!!]
   \centering
   \includegraphics [scale=0.7]
   {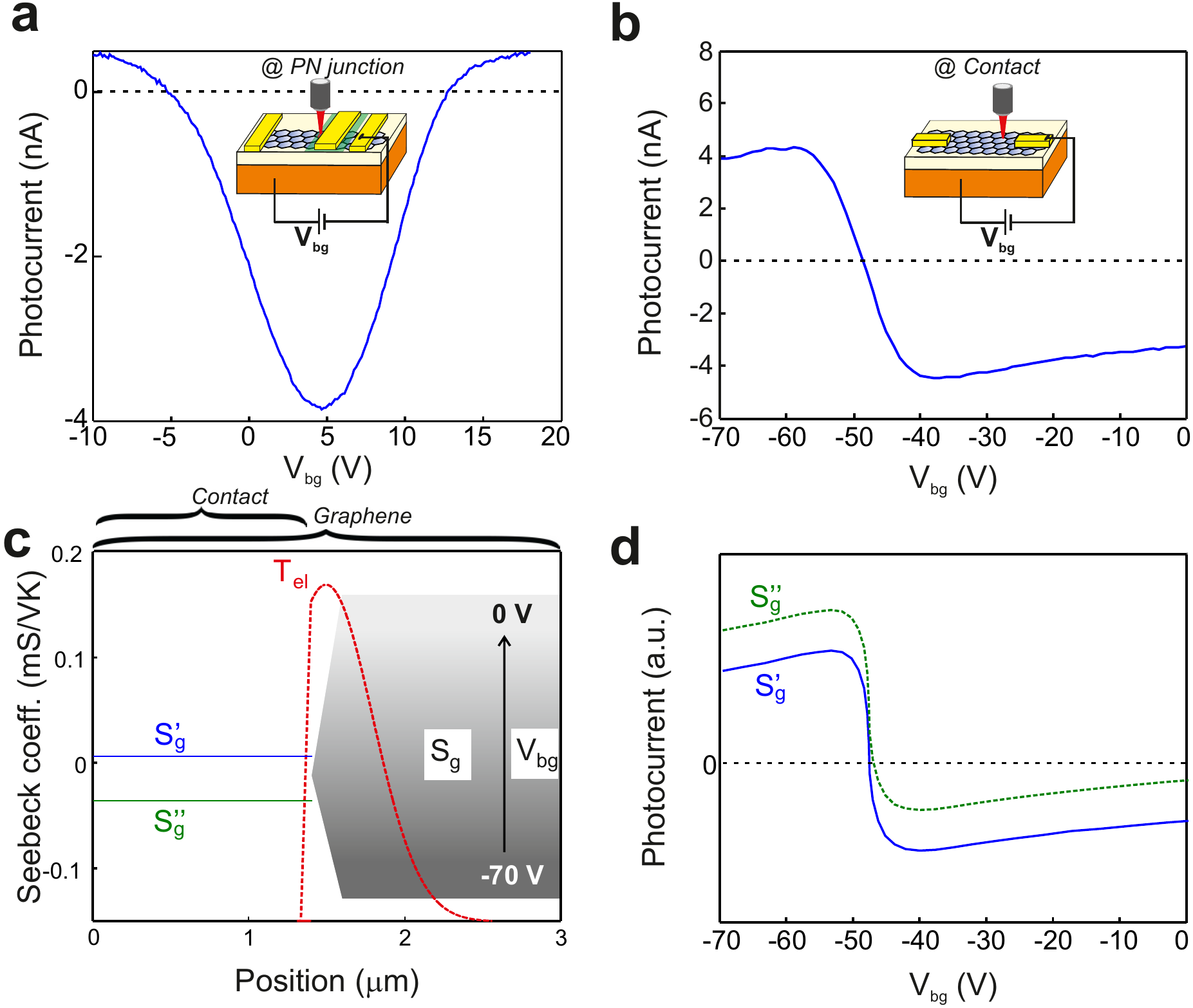}
\caption{\textit{Effect of Fermi energy on photocurrent. }
\textbf{a)} Photocurrent as a function of back gate voltage $V_{\rm bg}$ for the \textit{dual-gated device}. The multiple sign crossings and asymmetric gate effect clearly indicate PTE photocurrent generation \cite{Gabor2011,Song2011,Lemme2011}. The inset shows the measurement and device configuration, where the laser is fixed at 630 nm and the graphene carrier density is controlled by changing the back gate voltage, while the top gate is fixed at 0.4 V.
\textbf{b)} Photocurrent as a function of $V_{\rm bg}$ for the \textit{globally gated device}. The response is symmetric with one sign reversal at the Dirac point.
\textbf{c)} Details of the numerical simulation of the PTE photocurrent as a function of $V_{\rm bg}$. There are three regions with different Seebeck coefficients: the graphene underneath the gold contact with $S_{\rm g}'$ or $S_{\rm g}''$, a transition region, and the graphene sheet with the gate-dependent $S_{\rm g}$. The red dashed line represents the extent of the laser pulse and thus the region where the hot electrons are generated.
\textbf{d)} Simulation results for the gate-dependent PTE photocurrent using the Seebeck coefficient profile and hot electron profile as in \textbf{c}, for the two distinct Seebeck coefficients for the graphene underneath the metal contact.
}
    \label{Fig_GateDep}
\end{figure}

\clearpage

\begin{figure} [h!!!!!]
   \centering
   \includegraphics [scale=0.7]
   {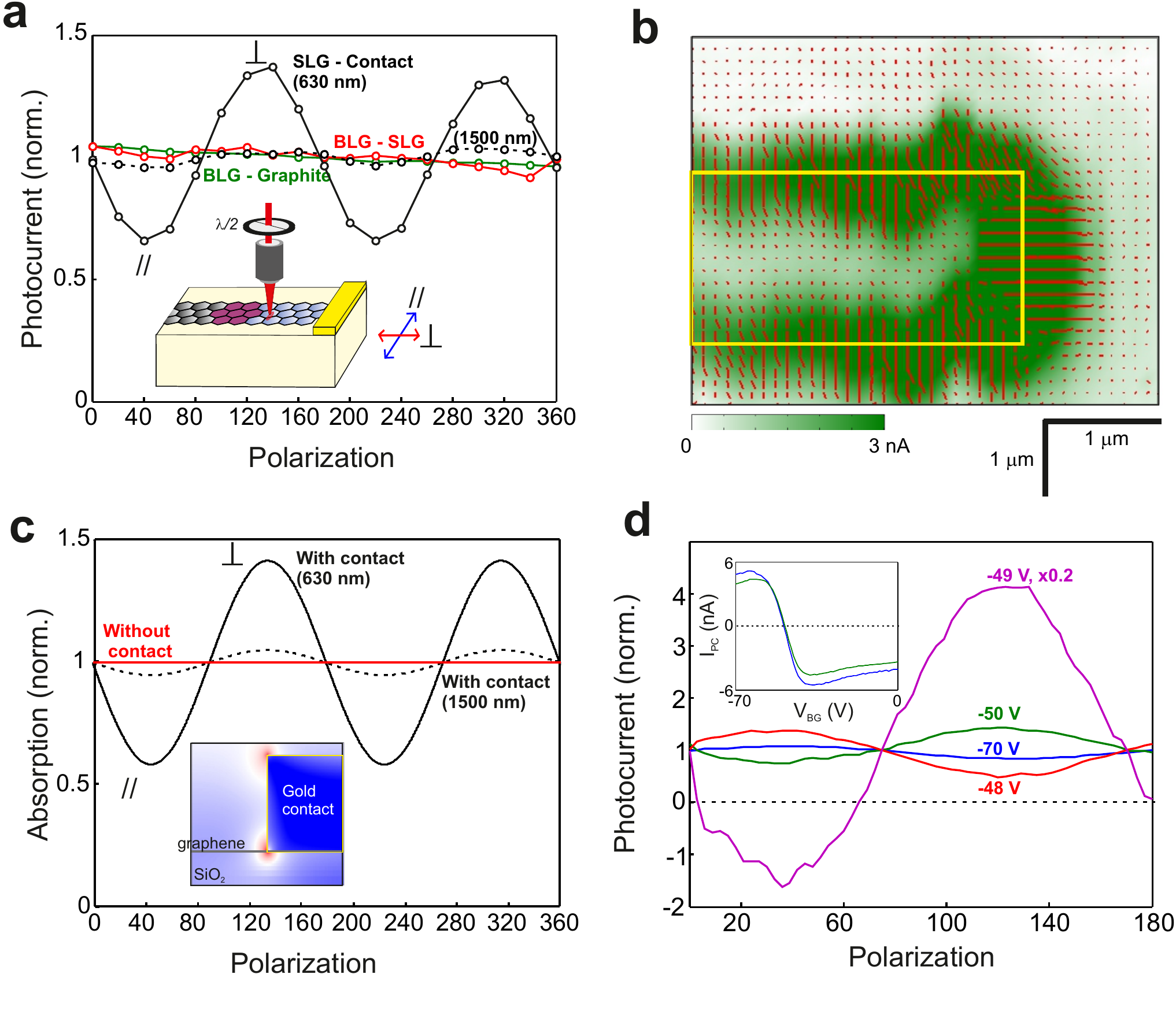}
\caption{\textit{Effect of light polarization on photocurrent. }
\textbf{a)} Photocurrent as a function of polarization for the \textit{transparent substrate device}. When light is focused at the BLG-SLG interface (red line) and the BLG-graphite interface (green line), there is no polarization dependence, whereas there is a strong polarization dependence for 630 nm excitation at the contact-graphene interface (black solid line) with increased (decreased) signal for light perpendicular (parallel) to the metal contact edge. Excitation with 1500 nm (black dashed line) leads to a lower contrast. The inset shows the device and measurement configuration, where a half wave plate is used to change the polarization of the incident light.
\textbf{b)} Photocurrent scanning microscopy image (green color scale) for the \textit{globally gated device}. The yellow line indicates the position of the metal contact. The red lines indicate the direction of maximum photocurrent, while their length indicates the magnitude of the photocurrent. The photocurrent is enhanced when the light is polarized perpendicular to the metal contact edge.
\textbf{c)} Results of numerical simulations (using Lumerical FDTD Solutions software), showing the polarization-dependent graphene absorption without metal contact (red line) and with metal contact for 630 nm excitation (black solid line) and 1500 nm excitation (black dashed line). The inset shows a side view of the field confinement that leads to the absorption enhancement for polarization perpendicular to the metal contact edge.
\textbf{d)} Photocurrent as a function of polarization for the \textit{globally gated device} for a number of different gate voltages, and an excitation wavelength of 630 nm. At this wavelength, there is photocurrent both due to direct, polarization-dependent graphene absorption and due to indirect, polarization-independent gold absorption. This leads to polarization-induced photocurrent sign reversal at gate voltage of -49 V. The inset shows the gate-dependent photocurrent for parallel (green) and perpendicular (blue) polarization.
}
    \label{Fig_PolDep}
\end{figure}

\clearpage

\end{document}